\documentclass[fleqn,usenatbib]{mnras}

\newcommand{\au}{\,\mathrm{AU}}

\newcommand{\yr}{\,\mathrm{yr}}
\newcommand{\kyr}{\,\mathrm{kyr}}

\newcommand{\pc}{\,\mathrm{pc}}

\def\lesssim{\mathrel{\hbox{\rlap{\hbox{\lower3pt\hbox{$\sim$}}}\hbox{\raise2pt\hbox{$<$}}}}}
\def\lesseq{\mathrel{\hbox{\rlap{\hbox{\lower3pt\hbox{$-$}}}\hbox{\raise2pt\hbox{$<$}}}}}
\def\gtrsim{\mathrel{\hbox{\rlap{\hbox{\lower3pt\hbox{$\sim$}}}\hbox{\raise2pt\hbox{$>$}}}}}
\def\gtreq{\mathrel{\hbox{\rlap{\hbox{\lower3pt\hbox{$-$}}}\hbox{\raise2pt\hbox{$>$}}}}}

\usepackage{newtxtext,newtxmath}

\usepackage[T1]{fontenc}

\DeclareRobustCommand{\VAN}[3]{#2}
\let\VANthebibliography\thebibliography
\def\thebibliography{\DeclareRobustCommand{\VAN}[3]{##3}\VANthebibliography}

\usepackage{graphicx}	
\usepackage{amsmath}	
\usepackage{cleveref}
\usepackage{hyperref}
\usepackage{natbib}
\usepackage{dblfloatfix}

\title[Accretion suppression in young eccentric binaries]{The protostar FU Orionis may not be bursting: dramatic accretion suppression in embedded young eccentric binaries}

\author[Rajika L. Kuruwita]{
Rajika L. Kuruwita,$^{1}$\thanks{E-mail: rajika.kuruwita@anu.edu.au}
\\
$^{1}$Heidelberg Institute for Theoretical Studies, Schloss-Wolfsbrunnenweg 35, 69118 Heidelberg, Germany\\
$^{2}$Research School of Astronomy \& Astrophysics, Australian National University, Cotter Rd, Canberra, 2611, Australia\\
}

\date{Accepted XXX. Received 30/07/2026; in original form ZZZ}

\pubyear{\the\year{}}

\begin{document}
\label{firstpage}
\pagerange{\pageref{firstpage}--\pageref{lastpage}}
\maketitle

\begin{abstract}
In 1936, the protostar FU Orionis began to brighten. Two years later, it was 100 times brighter and has not dimmed since. This object defined the class of FUor-type accretion bursts in protostars. Previous attempts to model FUors simulate isolated protostellar discs, thereby limiting the mass available to drive the burst and shortening its duration. However, these protostars are embedded within the interstellar medium, from which they actively accrete via accretion streamers. This work seeks the origin of FUors starting at interstellar medium scales. An unbiased search of star particles from a 25AU resolution star cluster formation simulation was performed to find FUor candidates. Most candidates were low-mass secondary stars in highly eccentric binaries, like FU Orionis. Zoom-in simulations down to 0.79AU resolution found candidates experience dramatic accretion suppression and bursts at the closest binary separation at periastron. Approaching periastron, the candidate’s orbital speed becomes supersonic striping away nearby gas. After periastron, the orbital speed decreases allowing gas to accrete onto the candidate rapidly via Bondi-Hoyle-Lyttleton (BHL) accretion. BHL accretion tails form behind the candidate and appear as accretion streamers. This dramatic suppression and burst near periastron is a physical mechanism for strong variable accretion in embedded young eccentric binaries. This mechanism also implies that FU Orionis returned to its former accretion rate.
\end{abstract}




\section{Introduction}


Accretion bursts in protostars were proposed as a solution to the `luminosity problem', which finds that protostars are dimmer than expected assuming a steady declining accretion rate \citep{kenyon_iras_1990}. Accretion bursts have been observed in protostars as short-term bursts for months to years in EXors \citep{herbig_1993-1994_2001}, or days to weeks in short-period protobinaries \citep{tofflemire_pulsed_2017}. FUor-type bursts \citep{wachmann_bisherige_1954, herbig_eruptive_1977} are also believed to be very long-duration bursts lasting decades to centuries.

FUor-type bursts are characterised by 1. a rapid rise in luminosity by 100-1000 times over a $1-20\yr$ period, and 2. a sustained peak luminosity for decades to centuries \citep{hartmann_fu_1996}. FUors also have distinctive reflection nebulae \citep{kenyon_fu_1995} because they are embedded in gaseous regions, and the brightness of the burst reflects off surrounding gas. Only a few dozen FUor objects where the burst is observed are known \citep{connelley_near-infrared_2018, guo_spectroscopic_2024}, and none have returned to their pre-burst luminosities. 

Proposed FUor mechanisms include gravitational instability in the protostellar disc \citep{kley_evolution_1999, armitage_episodic_2001, zhu_two-dimensional_2009}, mass pile-up in the disc due to magnetism \citep{konigl_effects_2011, dangelo_accretion_2012}, or tidal interaction with another star \citep{bonnell_binary_1992, reipurth_fu_2004, cuello_dust_2018}. FU Orionis itself was found to the lower-mass secondary star \citep{beck_nature_2012} of a binary \citep{wang_fu_2004}. Other FUors have been found to be in binaries \citep{green_testing_2016}; however, FUors are so bright that detecting nearby companions is difficult. Therefore, it is not possible to conclude if most FUors are in binary star systems.

The proposed mechanisms have struggled to reproduce both the rapid rise time and the sustained brightness. Previous attempts to model these events simulate isolated protostellar discs \citep{borchert_rise_2022, vorobyov_distinguishing_2021}, thereby limiting the mass available to drive the burst to the disc and shortening the burst duration. However, FUors are embedded within the interstellar medium, from which they actively accrete via accretion streamers. 

Accretion streamers carry gas from the interstellar medium down to protostar scales. Accretion streamers have been found in simulations for some time \citep{kuruwita_binary_2017, jorgensen_binarity_2022}; however, observations have only recently detected these features \citep{alves_gas_2019, luo_alma_2023}. Accretion streamers replenish the mass of protostellar discs and may trigger accretion bursts. Accretion streamers have been found to be synonymous with reflection nebulae \citep{gupta_reflections_2023}, suggesting that FUors are being fed by accretion streamers. This is supported by the recent discovery of an accretion streamer falling onto FU Orionis itself \citep{hales_discovery_2024}. The observational evidence emphasises the need to account for the surrounding ISM when modelling FUor-type bursts.

Here, we present the first work looking for the origin of FUor-type bursts starting from the interstellar medium.

\section{Methods}

\subsection*{MHD simulations of clustered star formation}

The cluster and zoom-in simulations \citep{kuffmeier_zoom-simulations_2017} were run with the magnetohydrodynamic code \texttt{RAMSES} \cite{teyssier_cosmological_2002}. The simulation of the clustered start formation setup is a $3000\,M_\odot$ piece of a molecular cloud in a 4~parsec cubic periodic box. The gas is modelled with an isothermal equation of state, and is initially drive solenoidal turbulence resulting in a typical velocity dispersion of 2~km~s$^{-1}$. The simulation is identical to that described in detail in \citet{haugbolle_stellar_2018} and \citet{kuruwita_contribution_2023}, but at higher resolution, reaching $25\au$ globally. 

The simulations use a standard sink particle prescription for formation. The star particle data were written out with a $2\yr$ cadence and coupled \citep{jensen_explaining_2018} with pre-stellar evolution models \citep{baraffe_new_2015} in \texttt{MESA} \citep{paxton_modules_2011} to derive protostar and accretion luminosities for all particles.

The gravitational softening radius is set to a third of the cell resolution, which in this simulation is $r_{soft}=7.5\au$. The accretion radius is defined as 4 grid cells, therefore in this cluster simulation the sink particles have an accretion radius of $100\au$. The prescription for mass accretion from a cell that is within the accretion radius of multiple sinks is that the mass is simply accreted onto the closest sink particle. This may introduce errors in edge cases where the closest sink particle is not the one that the gas in the cell is most bound to.

\begin{figure}\centering
  \centerline{\includegraphics[width=\linewidth]{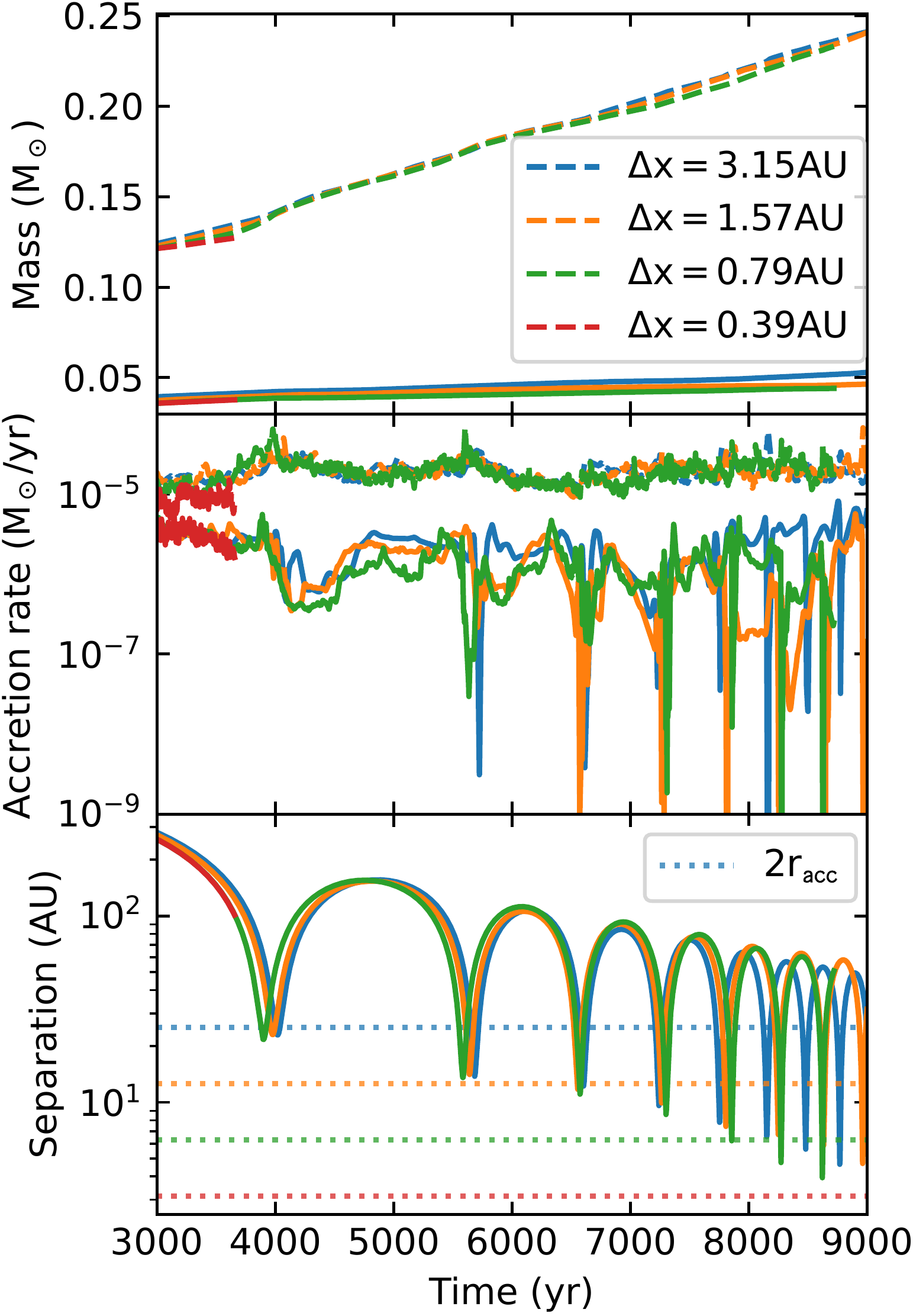}}
  \caption{{\bf Resolution study of the Candidate 3 zoom-in simulations}. \emph{Top:} Mass evolution of Candidates 3 (solid) and Primary star (dashed). The legend indicates the simulation resolution. \emph{Middle:} Accretion rate of the Primary (upper lines) and Candidate 3 (lower lines). \emph{Bottom:} Separation of the binary. The horizontal dashed lines annotate $2r_\mathrm{acc}$ at the resolution.}
  \label{fig:resolution_study}
\end{figure}

\subsection{FUor candidate identification from turbulent ISM simulation}


FUor candidates were identified by cross-correlating a step function with the luminosity profiles of star particles. To prepare the star particle data for correlation, a window of $100\yr$ is moved along the luminosity profile of the star. For each time window, if the variation in luminosity exceeded at least an order of magnitude ($\Delta L_{star}>1$) or the absolute magnitude variation was greater than 2.5, the luminosity profile within the window is scaled to range from $[0, 1]$. The scaled luminosity profile is then matched with the step function using \texttt{numpy} correlate function. The step function was defined as:

\begin{equation}
Y = \begin{cases}
 0 \text{\,\,\,for\,\,\,} 0\yr < x < 20\yr,\\
 1 \text{\,\,\,for\,\,\,} 20\yr < x < 100\yr.\\
\end{cases}
\end{equation}

If a median correlation of $>66.6$ is found, the event is saved for further inspection. 41 sink particles produced a match with this criteria, but after visual inspection of the matched events, 37 candidates remained. Four candidates were removed because the cross-correlated match did not show the characteristic rapid increase in accretion rate, but rather a smooth increase in the $100\yr$ time curve. 

Of the 37 candidates, most sink particles that had a match for a bursting event always had multiple matches. Upon further inspection of the long-term evolution of the candidates, 33 candidates were found to be binary or multiple star systems, while four showed very sporadic accretion history with multiple flybys. Eight candidates were found to be in binary star systems with another candidate, reducing these candidates to four binaries. 

Inspecting the long-term evolution of the 37 FUor candidates found that 33 were in binary star systems, while 4 experienced sporadic evolution from having interactions with multiple stars. Eight candidates were found to be in binary star systems with another candidate, reducing these candidates to four binaries. 

\begin{figure*}\centering
  \centerline{\includegraphics[width=\linewidth]{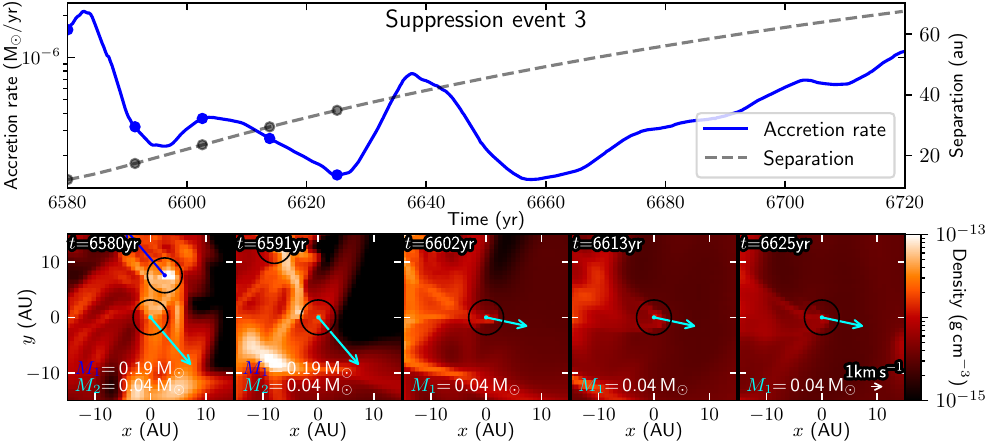}}
  \caption{{\bf Density projections of accretion suppression for the third periastron}. The top panel shows the accretion rate of Candidate 3 and the separation of the binary stars in blue solid and grey dashed lines, respectively. The circles on the top panel show where the density projections in the bottom panels were taken. The star particle arrows annotates the relative velocity between the stars, and the velocity legend refers to this. The black circles indicate the accretion radius of the star particles.}
  \label{fig:suppression_frames}
\end{figure*}

For the candidates in binary star systems, the strongest accretion variability is seen when the binary stars are at their closest separation at periastron. Previous studies of moderate-eccentricity proto-binaries show bursts of accretion onto the primary star can be triggered at periastron because the secondary star perturbs the circum-primary disc \citep{munoz_circumbinary_2020, kuruwita_dependence_2020}. The long-term evolution analysis found that as the binary becomes more eccentric and the periastron separation shrinks, the candidate luminosity suddenly drops and rises again. The cross-correlation FUor finding algorithm matched the step function to these dramatic accretion suppression events at periastron.

The orbital characteristics of the FUor candidates in binary stars show that the majority of candidates are the secondary star of the binaries with moderate to low mass ratios. Most event matches are found when the binary stars have in-spiralled to a semi-major axis of 10s of AU, and have moderate to high eccentricities ($0.5<e<1.0$). Some events were found in stars with hyperbolic trajectories with respect to the closest star, which may be flybys.

It should be noted that the separation at periastron of matched events is significantly below the accretion radius of the star particles (4 cell $=100\au$) and approaches the gravitational softening radius ($7.5\au$). It is possible that the observed suppression events are due to numerical effects from overlapping star-particle accretion radii. To determine whether the suppressed accretion is physical or numerical, zoom-in simulations are necessary. Due to limited time and computational resources, only zoom-in simulations on five candidates ran successfully. Because the cadence of the clustered star formation simulation is $10\kyr$, the time between the cluster simulation data dump and the candidate formation limits which zoom-in simulations successfully run.

\subsection{Zoom-in simulations on FUor candidates}

After candidate FUor star particles were identified, zoom-in simulations were performed on the candidates. Details of the method are described in \citep{kuffmeier_zoom-simulations_2017}. The data dump before the candidate forms is found from the cluster simulation and used to initialise the zoom-in simulations. The zoom-in is centred around where the candidate is expected to form, and the region within $10^4\au$ of the centre is allowed to experience full adaptive mesh refinement. Beyond this shell, the resolution is forced to a cell size $6446\au$. The computation domain is $4\pc^3$ like the cluster simulation, but high resolution is only applied around the candidate. This allows for the simultaneous low resolution modelling of the larger cloud while modelling at high resolution the region around the candidate. This allows for the movement from gas from larger scales down to the candidate via accretion streams. 

The zoom-in simulations use a piecewise polytropic equation of state given by described by \citet{masunaga_radiation_2000}. These values describe the gas behaviour during the initial isothermal collapse of the molecular core, adiabatic heating of the first core, the H2 dissociation during the second collapse into the second core, and the return to adiabatic heating.

Due to limited computational resources and time, not all simulations progressed to the formation of the candidate. Factors such as the time difference between the cluster simulation data dump and the candidate formation, or the mean density, impact how soon the zoom-in simulation will create the sink particle. These other simulations are continuing to run and will be presented in future publications. 

The initial highest level of refinement for the zoom-in simulations is $3.15\au$, and five simulations run until the formation of the candidate. The initial evolution of the zoom-in simulations found that some systems still show suppressed accretion, while others do not. This was dependent on the eccentricity and the separation at periastron.

Resolution studies are attempted on all simulations with subsequent zoom-ins pushing the simulations to double the previous resolution. The zoom-in on Candidate 3 completed a resolution study with the highest level of refinement until $0.79\au$. A zoom-in with resolution $0.39\au$ was also run, but this simulation would not complete a few binary orbits in a reasonable time. The evolution of the binaries from the zoom-in simulations is shown in \Cref{fig:resolution_study}.

We are certain that the accretion rate measured is not affected by numerical issues from overlapping accretion radii while the separation is larger than twice the accretion radius of the star particles. This means that the $\Delta x=0.79\au$ Candidate 3 simulation accurately resolves the accretion evolution of the star particles for the first five periastrons. 

\begin{figure*}\centering
  \centerline{\includegraphics[width=\linewidth]{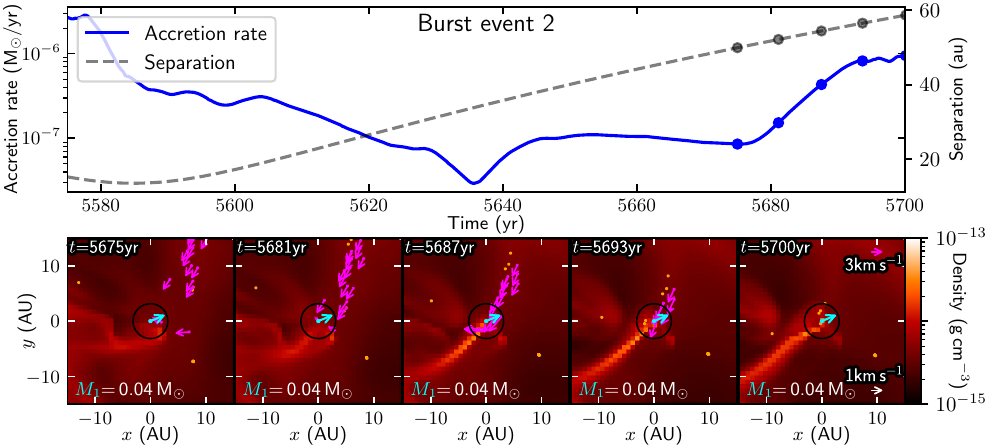}}
  \caption{{\bf Density projections of accretion burst for the second periastron}. Same as \Cref{fig:suppression_frames}. The magenta points and arrows show tracer particles that are accreted over the burst, and velocity legend on the upper right of the last panel refers this. The yellow points are other tracer particles that are accreted soon after by the candidate.}
  \label{fig:burst_frames}
\end{figure*}

\begin{figure*}\centering
  \centerline{\includegraphics[width=\linewidth]{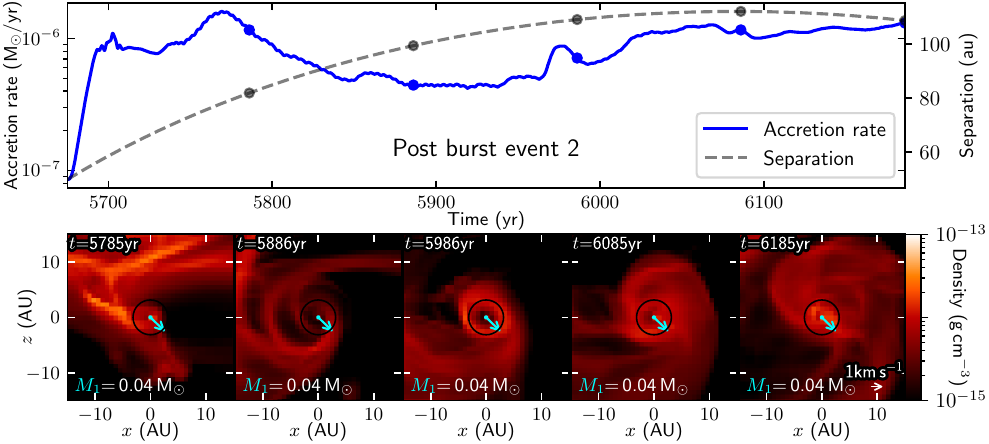}}
  \caption{{\bf Density projections of the gas around Candidate 3 post-accretion burst}. Same as \Cref{fig:suppression_frames}.}
  \label{fig:post_burst_frames}
\end{figure*}

In the second periastron onward when the periastron separation falls below $20\au$, the dramatic accretion suppression and bursts are observed, as seen in the long-term evolution plots. The depth of accretion suppression between events varies due to the turbulent environment, leading to stochastic behaviour. To understand the mechanisms causing the suppression and bursts, we examine each periastron in detail.

\section{Results}

\subsection{Accretion suppression mechanism}

As the candidate approaches periastron, its orbital speed significantly increases, becoming supersonic relative to the medium. This results in a characteristic bow shock around the candidate, as seen in \Cref{fig:suppression_frames}. The increasing orbital speed strips material away from the candidate, shutting off accretion soon after periastron.

\subsection{Accretion burst mechanism}

Post-periastron, the candidate's orbital speed decreases, allowing accretion to resume via Bondi-Hoyle-Lyttleton (BHL) accretion \citep{bondi_mechanism_1944, hoyle_accretion_1941}. This results in the characteristic BHL accretion tail forming behind the candidate as it leaves periastron, as seen in \Cref{fig:burst_frames}. The accreted mass approaches the candidate predominantly from the direction of travel. There is some curvature in the accreted mass trajectory due to the binary orbital motion. The gravity of the candidate bends gas trajectories behind the candidate, creating a dense tail. This also reduces the relative speed between the gas and the candidate, allowing it to be accreted. This BHL tail contains infalling material and may be observed as an accretion streamer.

\subsection{Post-accretion burst evolution}

Post-accretion burst, the BHL tail remains until the candidate's orbital speed has slowed down sufficiently. At this point, the gas in the BHL tail can `catch up' to the candidate and spirals around it, forming disc-like structures, as seen in \Cref{fig:post_burst_frames}. For a couple of centuries after the burst, hints of the BHL tail persist as an accretion streamer that feeds the candidate.

\section{Conclusions}


The results of this work found that protostars can produce FUor-like accretion profiles after dramatic accretion suppression at periastron in young eccentric binaries. Zoom-in simulations resolved the periastron passage in an FUor candidate, confirming that this accretion suppression is a physical mechanism. The accretion suppression and burst occur as a natural consequence of the orbital speed variation in highly eccentric binaries and accretion from an ambient medium. It would not be possible to model this accretion variation in young stars without accounting for accretion from the interstellar medium.


While the orbital speed of the candidate is low relative to the surrounding medium, accretion is driven by streamers. As the orbital speed increases approaching periastron, the candidate speed becomes supersonic relative to the surrounding medium. This results in a bow shock forming, which strips away nearby gas, suppressing accretion from the medium. Post-periastron, the orbital speed of the candidate decreases and accretion resumes via Bondi-Hoyle-Lyttleton (BHL) accretion, forming a dense tail. As the orbital speed continues to decrease, the BHL tail can catch up to the candidate and form a disc around it. Accretion from other streamers is also able to form due to the low relative speed. 

While the zoom-in simulation resolves only the first few orbits before the periastron separation becomes smaller than two accretion radii, we are confident that the accretion suppression seen in later orbits is not due to numerical effects from overlapping accretion radii. From the zoom-in simulations, we are also confident that the accretion suppression seen in the long-term candidate evolution is also physical as long as the binary orbit remains highly eccentric and accretes from the surrounding medium. This effect will be enhanced in highly eccentric binaries with low mass-ratio binaries, as the secondary star will have even stronger variation in orbital speed near periastron. This behaviour should diminish as the binary evolves and is no longer embedded in its star-forming cloud.

The mass of the zoom-in candidate during the resolved orbits is significantly lower than FU Orionis, because the candidate has just formed. However, the dominant driver of the accretion variation is the orbital speed, and we expect the candidate mass to play a smaller role. The BHL tails formed will generally point in the direction of the other star in the binary. The accretion streamer found around FU Orionis is also pointing in the direction of its companion \citep{hales_discovery_2024}. Based on this mechanism, we hypothesise that this accretion streamer may be a remnant of the BHL accretion tail, and the burst observed 90 years ago is the star returning to its pre-suppression accretion rate.

\vspace{-0.5cm}
\section*{Acknowledgements}
RLK Acknowledges support from the Klaus Tschira Foundation through the Heidelberg Institute for Theoretical Studies (HITS) and their Independent Postdoc Program. RLK receives funding from the Australian Research Council (Discovery Project grants DP230102280 and DP250101526). RLK acknowledges high performance computing resources provided by the University of Copenhagen and Australian National Computational Infrastructure (grant ek9) and the Pawsey Supercomputing Centre (project pawsey0810) in the framework of the National Computational Merit Allocation Scheme and the ANU Merit Allocation Scheme. RLK thanks Prof. Daniel Price (Monash University) and Prof. Christoph Federrath (Austrlaian National University) for their support and insightful discussion. yt \citep{turk_yt:_2011} was used to help visualise and analyse these simulations.

\section*{Data Availability}

The MHD and gravity solvers used in this study are closely related to the public RAMSES version available at \url{https://bitbucket.org/rteyssie/ramses}. The numerical methods, developed in Copenhagen, relevant for star formation are described in Haugbølle et al (2018) \citep{haugbolle_stellar_2018}. Original simulation data generated and/or analysed during the current study are available from the corresponding author upon reasonable request.



\bibliographystyle{mnras}
\bibliography{Bibliography} 

\bsp	
\label{lastpage}
\end{document}